\documentclass[preprint,onecolumn,nofootinbib]{revtex4}

\usepackage[colorlinks=true,linkcolor=blue,urlcolor=blue,filecolor=black,citecolor=red,
            pdfstartview=FitV,pdftitle={},pdfsubject={},pdfkeywords={},pdfpagemode=None,
            bookmarksopen=true]{hyperref}
\usepackage{graphicx}   
\usepackage{amsmath}
\usepackage{amsfonts}
\usepackage{amssymb,ulem}
\usepackage{color}
\usepackage{tikz}
\usepackage{dcolumn}    
\usepackage{float}
\usepackage{xcolor}
\usepackage{subfig}
\usepackage{overpic}

\begin{document}
\title{
	Holographic Entanglement Entropy as a Probe of Dynamical Criticality in Scalarizing Black Holes
}

\author{Yi Li $^{1}$}
\email{yilyu@stu2022.jnu.edu.cn}

\author{Ke-tai Wu $^{1}$}
\email{tankwu888@gmail.com}

\author{Chong-Ye Chen $^{1}$}
\email{cychen@stu2022.jnu.edu.cn}

\author{Chao Niu $^{1}$}
\email{niuchaophy@gmail.com}

\author{Cheng-Yong Zhang $^{1}$}
\email{zhangcy@email.jnu.edu.cn}

\author{Peng Liu $^{1}$}
\email{phylp@email.jnu.edu.cn}
\thanks{Corresponding author}

\affiliation{
	$^1$ Department of Physics, Jinan University, Guangzhou 510632, China
}

\begin{abstract}

	We demonstrate that holographic entanglement entropy (HEE) serves as a powerful diagnostic tool for both static and dynamical critical phenomena in the Einstein-Born-Infeld-Scalar (EBIS) model. While HEE is well-known for capturing static phase transitions, we reveal its novel ability to probe dynamical criticality, particularly the ``flip'' phenomenon-a sign inversion in the scalar field at a critical point. Near the flip, HEE exhibits relaxation dynamics that closely mirror those of the scalar field, with both relaxation times scaling logarithmically with the distance from the critical point. This intimate connection between the relaxation of HEE and the scalar field highlights HEE as a sensitive probe of dynamical critical phenomena. Our findings provide new insights into the interplay between quantum information and gravitational dynamics, offering a deeper understanding of critical behavior in strongly coupled systems.
	
\end{abstract}

\maketitle
\tableofcontents

\section{Introduction}
\vspace{10pt}
\label{sec:introduction}

In recent years, the study of critical phenomenon in holographic models based on AdS/CFT correspondence has attracted significant attention, particularly in the context of strong coupling systems like condensed matter theory (CMT) and quantum chromodynamics (QCD) \cite{Gubser:2018cha,Gubser:2017pyx,ling2016novel,ling2016characterization,ling2016holographic,ling2015building,donos2014holographic,donos2013interaction,donos2014holoqlattices,donos2014novel,kiritsis2015holographic,baggioli2015electron,Gong:2023tbg,Liu:2023rhd,Chowdhury:2024lqw,Barbosa:2024pyn,Liu:2020blk,Yang:2023wuw,Liu:2021rks}. These extensive research focus on the static critical phenomena for the black hole system in the stable state, such as thermal critical phenomena and quantum critical phenomena. The holographic entanglement entropy (HEE) has emerged as a powerful tool for identifying critical points in these critical phenomena, providing insights into the underlying interplay between quantum information and critical phenomena \cite{ling2016characterization,ling2016holographic,Liu:2023rhd,Chowdhury:2024lqw,Barbosa:2024pyn,Yang:2023wuw,Liu:2021rks,Liu:2020blk}. On the other hand, as an important phenomenon in CMT, dynamical critical phenomena (DCP) are distinct from static critical phenomena, as they involve the evolution of a system, often exhibiting complex and nontrivial behavior \cite{Heyl:2013ywe,Meibohm:2022,Canovi:2014,Eckstein:2009,Garrahan:2010,Diehl:2010}. However, DCP have not been fully explored due to the complexity of the dynamical process in strongly coupled systems.

In the context of holography, critical phenomena have been observed in various models, such as the Einstein-Maxwell-scalar (EMS) model, the Einstein-Born-Infeld-scalar (EBIS) model, etc \cite{Wu:2024,Zhang:2022cmu,Chen:2023iws,Niu:2022zlf,Mou:2024ecv,Cai:2004eh,Zhang:2021kha,Zhang:2020obn,Bi:2020vcg}. These models exhibit rich critical phenomena, including the emergence of critical points and the presence of dynamical criticality. In the time-dependent holographic models, the covariant holographic entanglement entropy has been proposed as a useful measure for studying the dynamics of the systems \cite{Hubeny:2007xt,Ling:2019tbi,Liu:2015lit,Chou:2023adi,Ecker:2018jgh,Kaplan:2022orm,Jeong:2023lkc,Wang:2019ued,Hubeny:2018ijt,Grado-White:2020jqn,Colafranceschi:2020dxk,Auzzi:2019mah}. The covariant HEE provides a geometric interpretation of the entanglement entropy, allowing us to probe the dynamics of entanglement in strongly coupled systems. As a result, the covariant HEE serves as a valuable diagnostic tool for identifying dynamical critical phenomena, reflecting the underlying physics of quantum entanglement and information in strongly coupled systems \cite{Paz:2024clg}.

Our study first points out that HEE can capture the static critical phenomenon between scalarized phase (scalarized black hole) and normal phase (AdS-BI black hole) in the EBIS model. Furthermore, this work investigates the dynamical critical phenomenon denoted as flip using covariant HEE. Specifically, the covariant HEE exhibits the critical relaxation phenomenon at the flip point, which is specifically reflected in the dynamical critical behavior and the quantum information of the system.

This paper is organized as follows: In Section \ref{sec:model_description}, we will provide a detailed description about the setup of the EBIS model, including the action and equations of motion. Section \ref{sec:result_time_evolution} will review the time evolution of scalar fields in the EBIS model, analyzing the emergence of dynamical critical phenomena, particularly the observed flip phenomenon. In Section \ref{sec:hee_intro}, we will introduce covariant HEE. In the following Section \ref{sec:hee_static}, we will examine how the HEE of the evolved final state varies with changes in system parameters, investigating the influence of scalar hair on the system, proving HEE with the final state can capture the critical phenomenon between hairy and hairless black hole configurations. Finally, in Section \ref{sec:hee_dynamic}, we will discuss the dynamical evolution process of HEE, emphasizing the role of HEE as a probe for capturing the dynamics.

\section{Einstein-Born-Infeld-scalar model}
\vspace{10pt}
\label{sec:model_description}

The action of the EBIS theory reads
\begin{equation}\label{eq:action}
	S = \int d^{4}x \sqrt{-g} \left[ R - 2\Lambda - 2\partial_{\mu}\phi\partial^{\mu}\phi + \frac{4f(\phi)}{a}\left( 1 - \sqrt{1 + a \frac{F_{\mu\nu}F^{\mu\nu}}{2}} \right) \right].
\end{equation}
where $R$ is the Ricci scalar and $\Lambda$ is the cosmological constant. $a$ is the nonlinear correction parameter in Born-Infeld theory, which we call BI factor in the following context. When $ a \to 0 $, the action approaches the EMS limit, while as $ a \to \infty $ the electromagnetic field will disappear from the action, the action approaches the ES limit. The $2$-form field $F_{\mu\nu} = \partial_\mu A_{\nu} - \partial_\nu A_{\mu}$ represents the Born-Infeld electromagnetic field strength tensor, and $f(\phi)$ is the non-minimal coupling function between real scalar field $\phi$ and Maxwell field $A_{\mu}$. We take $16\pi G=1$ for simplicity. In this work, we focus on couplng function $ f(\phi) = e^{\beta\phi^2} $. By variating the action, we can obtain the equations of motion for this model.  
\begin{equation}
	\begin{aligned}
		R_{\mu\nu} - \frac{1}{2} R g_{\mu\nu} + \Lambda g_{\mu\nu}                                                           & = \frac{\tau_{\mu\nu}}{2},                                                            \\ 
		\partial_{\mu}\left[\frac{\sqrt{-g} f(\phi) F^{\mu\nu}}{\sqrt{1 + a \frac{F_{\rho\sigma} F^{\rho\sigma}}{2}}}\right] & = 0,                                                                                  \\ 
		\frac{\partial_{\mu}(\sqrt{-g}\partial^{\mu}\phi)}{\sqrt{-g}}                                                        & = -\dot{f}(\phi) \frac{1 - \sqrt{1 + a \frac{F_{\rho\sigma} F^{\rho\sigma}}{2}}}{a}.
	\end{aligned}
\end{equation}
where $\dot f(\phi) \equiv \frac{df(\phi)}{d\phi}$, and $\tau_{\mu\nu}$ represent the energy-momentum tensor, which can be expressed as
\begin{equation}
	\label{eq:emtensor}
	\begin{aligned}
		\tau_{\mu \nu} = 4\left({\partial_{\mu}\phi \partial_{\nu}\phi - \frac{1}{2}g_{\mu \nu} \partial_{\rho}\phi  \partial^{\rho}\phi}\right) {+ 4f(\phi)}\left[{\frac{1-\sqrt{1+a F_{\rho\sigma}F^{\rho\sigma}/2}}{a}g_{\mu \nu} + \frac{F_{\mu \rho} F_{\nu}{}^{\rho}}{\sqrt{1+a F_{\rho\sigma}F^{\rho\sigma}/2}}}\right]
	\end{aligned}
\end{equation}
We take the ingoing Eddington-Finkelstein coordinate here to explore the nonlinear dynamics of a spherical asymptotically AdS black hole. The metric in this coordinate reads
\begin{equation}
	\label{eq:ansatz}
	\begin{aligned}
		d s ^ { 2 } & = - \alpha(v,r) d v ^ { 2 } + 2 d v d r +\zeta(v,r)^{2}(d \theta ^{2}+\sin^{2}\theta d \varphi ^{2}), 
	\end{aligned}
\end{equation}
where $\alpha$, $\zeta$ are functions of $(v, r)$. For simplicity, we take the following auxiliary variables
\begin{align}
	S & \equiv \partial_v\zeta + \frac{1}{2}\alpha\partial_r\zeta, \label{eq:eqauxvarS} \\
	P & \equiv \partial_v\phi + \frac{1}{2}\alpha\partial_r\phi. \label{eq:eqauxvarSP}
\end{align}
At apparent horizon, $g^{\mu \nu}\partial_{\mu}{\zeta}\partial_{\nu}{\zeta} = 0 $, and accordingly the auxiliary variable should satisfy $S=0$. The BH irreducible mass $M_h \equiv \sqrt{\frac{V_h}{4 \pi}}=\zeta(r_h,v)$ where $V_h$ is the apparent horizon area, measuring the thermodynamic entropy of the BH. For scalar field and gauge field, we require $\phi=\phi(v,r)$ and $A_{\mu}dx^{\mu}=A(v,r)dv$. Due to the conservation law, $\partial_r A = \frac{Q}{\zeta^{2} f} $, where $Q$ is the BH charge. All these variables $(\alpha,\zeta,\phi,A_t,S,P)$ are functions of time $v$ and radial coordinate $r$. By substituting the auxiliary variables into Einstein field equations, we can obtain the following equations for the spatial derivatives of the variables
\begin{align}
	\partial_r^2 \zeta  & = -\zeta (\partial_r \phi)^2, \label{solve_z}                                                                                                                                                                                                                                      \\
	\partial_r S        & = \frac{1 - 2 S \partial_r \zeta}{2 \zeta} - \frac{\Lambda \zeta}{2} + \frac{f(\phi) \zeta \left(1 - \sqrt{\frac{f(\phi)^2\zeta^4}{a Q^2 + f(\phi)^2\zeta^4}}\right)}{a} - \frac{Q^2 \sqrt{\frac{f(\phi)^2 \zeta^4}{a Q^2 + f(\phi)^2 \zeta^4}}}{f(\phi) \zeta^3}, \label{solve_s} \\
	\partial_r^2 \alpha & = -4 P \partial_r \phi + \frac{4 S \partial_r \zeta - 2}{\zeta^2} + \frac{4 Q^2 \sqrt{\frac{f(\phi)^2 \zeta^4}{a Q^2 + f(\phi)^2 \zeta^4}}}{f(\phi) \zeta^4}. \label{solve_a}
\end{align}
And scalar equation gives
\begin{equation}
	\label{solve_P}
	\begin{aligned}
		\partial_r P = - \frac{P \partial_r \zeta + S \partial_r \phi}{\zeta} - \frac{1 - \sqrt{\frac{f(\phi)^2 \zeta^4}{a Q^2 + f(\phi)^2 \zeta^4}} }{2a} \frac{df}{d\phi}.
	\end{aligned}
\end{equation}
For time direction, we mainly focus on the evolution of the scalar field $\phi$ using the definition of the auxiliary variable $P$,
\begin{equation}
	\label{solve_phi}
	\partial_v \phi = P - \frac{1}{2} \partial_{r} \phi.
\end{equation}
On the other hand, the evolution equation of the auxiliary variable $S$
\begin{equation}
	\partial_v S = \frac{S \partial_r \alpha - \alpha \partial_r S}{2} - \zeta P^2
\end{equation}
is also needed to fix extra gauge freedom. 

To obtain the numerical solution, we employ a nested iterative scheme. At each time step, we first solve for the spatial configuration of the metric functions and auxiliary variables. This involves an subsequent steps where we successively solve Eq. \eqref{solve_z} for $\zeta(r)$, followed sequentially by Eq. \eqref{solve_s} for $S(r)$, Eq. \eqref{solve_a} for $\alpha(r)$, and finally Eq. \eqref{solve_P} for $P(r)$. After applying this method to determine the spatial profiles, we can evolve the scalar field $\phi$ in time using Eq. \eqref{solve_phi}. The entire process is then repeated for each time steps. This nested approach ensures self-consistency in the spatial solutions at each time step while advancing the system's evolution, enabling us to capture the fully nonlinear dynamics of scalar field and its interaction with the background spacetime.

The asymptotic behaviors of the variables are
\begin{equation}
	\label{asymptotic behaviors}
	\begin{aligned}
		\phi(v,r)   & = \frac{\phi_3 (v)}{r^3} + \frac{3}{8 \Lambda r^4} \left(\frac{Q^2 f'(0)}{f(0)^2} - 8 \phi_3 ' (v)\right) + O\left(r^{-5}\right),                                         \\
		\zeta(v,r)  & = r - \frac{3 \phi_3^2 (v)}{10 r^5} + \frac{3 \phi_3 (v)}{14 \Lambda r^6} \left(\frac{Q^2 f'(0)}{f(0)^2} - 8 \phi_3 ' (v)\right) + O\left(r^{-7}\right),                  \\
		S(v,r)      & = - \frac{\Lambda}{6} r^2 + \frac{1}{2} - \frac{M}{r} + \frac{Q^2}{2 f(0) r^2} - \frac{3 \Lambda}{20 r^4} \phi_3^2 (v) + O\left(r^{-5}\right),                            \\
		P(v,r)      & = \frac{\Lambda \phi_3 (v)}{2 r^2} + \frac{1}{r^3} \left(\frac{Q^2 f'(0)}{4 f(0)^2} -  \phi_3 ' (v)\right) + \frac{3}{2\Lambda r^4} \phi_3 '' (v) + O\left(r^{-5}\right), \\
		\alpha(v,r) & = - \frac{\Lambda}{3} r^2 + 1 - \frac{2M}{r} + \frac{Q^2}{f(0) r^2} + \frac{\Lambda}{5 r^4} \phi_3^2(v) + O\left(r^{-5}\right).
	\end{aligned}
\end{equation}
where $f'(\phi) = \frac{df(\phi)}{d\phi}\Big|_{r=0}$, and $\phi_3'(v) = \frac{d\phi_3(v)}{dv}\Big|_{r=0}$. The constant $M$ represents the Arnowitt-Deser-Misner (ADM) mass and $Q$ is the charge of the spacetime. To eliminate the singularities of these variables at the AdS boundary, we perform the variable redefinition:
\begin{equation}
	\label{eq:variable_substitutions}
	\begin{aligned}
		\phi(v,r)   & = \frac{\phi_1(r,v)}{r^3},                                        \\
		\zeta(v,r)  & = r + \lambda + \frac{\zeta_1(v,r)}{r^{3}},                       \\
		S(v,r)      & = \frac{(r+\lambda)^{2}}{2} + \frac{1}{2} - \frac{S_{1}(v,r)}{r}, \\
		P(v,r)      & = \frac{-3\phi_{1}(r)}{2r^{2}} + \frac{P_{1}(v,r)}{r^{3}},        \\
		\alpha(v,r) & = (r+\lambda)^{2} + 1 + \alpha_1(v,r).
	\end{aligned}
\end{equation}
Furthermore, to facilitate our numerical analysis more practicably, we introduce a gauge degree of freedom $\lambda$. This allows us to redefine the physical event horizon radius $r_{h,\text{phys}}$ in terms of a fixed horizon $r_h$ in a new radial coordinate using this time-dependent gauge $\lambda(v)$:
\begin{equation}
	\label{eq:physical_horizon}
	r_{h,\text{phys}}(v) = r_h + \lambda(v).
\end{equation}
Through this transformation, we can attribute variations in the physical horizon to evolution of $\lambda(v)$ while maintaining the horizon radius $r_h=1$. This approach significantly reduces the computational complexity while maintaining the physical accuracy of our solution. The evolution of $\lambda(v)$ thus captures the dynamics of the event horizon in a computationally efficient manner.

At each time slice, the value of $\lambda$ is determined by the boundary condition $S(r_h, \lambda) = 0$, and subsequently the evolution evolution of $\lambda$ reads \cite{Zhang:2022cmu}
\begin{equation}
	\label{partial_t}
	\partial_v\lambda \big|_{v=v_0} = -\frac{1}{2}\lim_{r\rightarrow\infty}[\alpha(v_0,r) - (r+\lambda)^2 - 1]=-\frac{1}{2}\lim_{r\rightarrow\infty}\alpha_{1}(v_0,r),
\end{equation}
Before the numerical simulation of time evolution, we set up the initial configuration of $\phi$ as a Gaussian wave packet. We then solve the nested spatial equations using the Chebyshev-Lobatto collocation method with $60$ points. This chosen number of points was determined to be optimal through convergence testing with finer grids.

The spatial discretization allows us to solve for the initial conditions of all variables. Once we have the initial state, we employ a fourth-order Runge-Kutta (RK4) method to evolve the system in time. The RK4 method provides a good balance between accuracy and computational efficiency for this type of system.

The process for each time step is as follows:
\begin{enumerate}
	\item Solve the spatial equations using the Chebyshev-Lobatto collocation method.
	\item Use the RK4 method to evolve the time-dependent variables ($\phi$, $P$, $S$, and $\lambda$).
	\item Update the boundary condition $S(r_h, \lambda) = 0$ to determine the new value of $\lambda$. And use the RK4 method to evolve the $\phi$ at next time step.
	\item Repeat steps 1-3 for each time step.
\end{enumerate}

\section{Dynamical Criticality and the Flip Phenomenon in Scalarizing Black Holes}
\label{sec:result_time_evolution}

In this section, we review key findings from our numerical simulations, beginning with the time evolution of scalar fields \cite{Wu:2024}. A crucial aspect of our discussion is the flip phenomenon, a significant dynamical critical phenomenon observed in this context. 

In previous studies, we used Gaussian wave packets as initial fluctuations to induce spontaneous scalarization in the system,
\begin{equation}
	\begin{aligned}
		\phi(v=0,r)=p e^{-\left(\frac{r-r_{0}}{w}\right)^{2}},
	\end{aligned}
\end{equation}
where $p$, $r_{0}$, and $w$ represent the initial amplitude, center, and width of the Gaussian wave packet, respectively. We set $w=r_h$ and $r_0=4r_h$, with $r_h=1$ for numerical simplicity, while the amplitude of initial wave packet $p$ represents the magnitude of the initial scalar field, remaining a key parameter in our study of spontaneous scalarization in EBIS model.

\begin{figure}[H]
	\centering
	\includegraphics[width=0.6\linewidth]{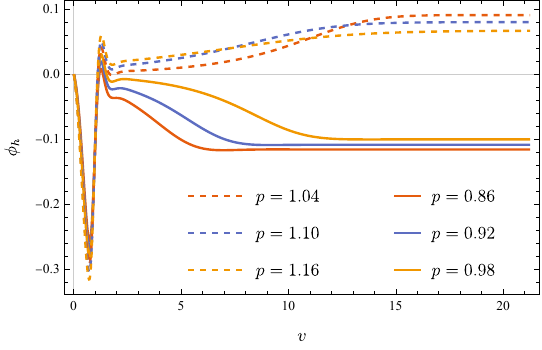}
	\caption{Evolution of $\phi_{h}$ over time, with initial data: $\phi_{0}=pe^{-64({0.5-\frac{1}{r}})^{2}}$ with $Q =0.6 $, $a=0.01$, $\beta=40$.} 
	\label{evolution_of_phi}
\end{figure}

When studying the full back-reaction gravitational system, we encounter intriguing phenomena, particularly in the behavior of the scalar field $ \phi $ around critical values of a parameter $ p $. For instance, the horizon value of $ \phi $, as illustrated in Fig. \ref{evolution_of_phi}, we observe a significant change in the final value of $ \phi_h $ as varying $p$ across the critical value $ p_s $. Specifically, for $ p < p_s $, $ \phi_h $ stabilizes at a negative value, while for $ p > p_s $, $ \phi_h $ stabilizes at a positive value, with the absolute magnitudes of $ \phi_h $ being nearly identical in both cases. This type of behavior of the scalar field can be seen across the entire radial axes $z$, suggesting that this criticality reflects a broader, underlying dynamical critical behavior in the system. The sharp change in the stabilized values around $ p_s $ indicates that the system may undergo a form of transition, with $ p_s $ marking the critical point where the qualitative nature of scalar field $ \phi $ shifts dramatically.

\begin{figure}[H]
	\centering
	\includegraphics[width=0.6\linewidth]{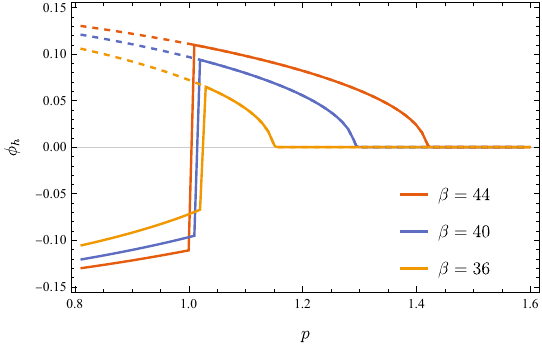}
	\caption{The solid line represents the final values of $\phi_{h}$ at the horizon, while the dashed line represents $|\phi_{h}|$ with initial data: $Q=0.6$, $a=0.01$. }
	\label{fig:fi}
\end{figure}

\begin{figure}[H]
	\centering
	\includegraphics[width=0.6\linewidth]{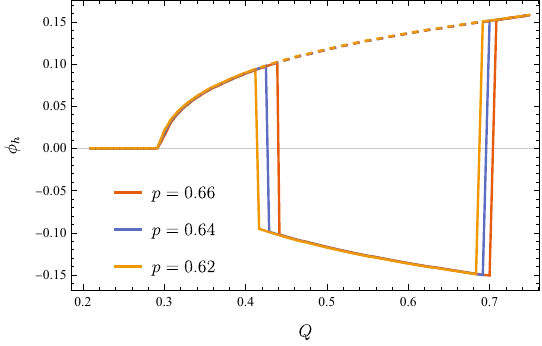}
	\caption{The solid line represents the final values of $\phi_{h}$ at the horizon, while the dashed line represents $|\phi_{h}|$ with initial data:  $\beta=150$, $a=0.01$. }
	\label{fig:Q}
\end{figure}

\begin{figure}[H]
	\centering
	\includegraphics[width=0.6\linewidth]{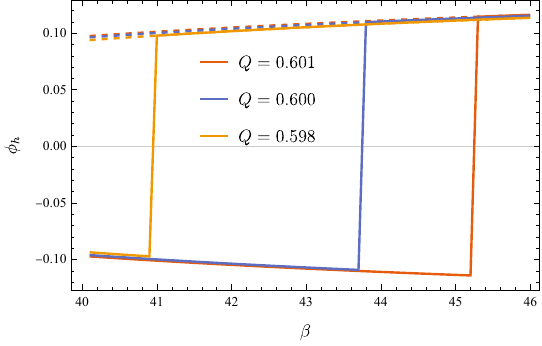}
	\caption{The solid line represents the final values of $\phi_{h}$ at the horizon, while the dashed line represents $|\phi_{h}|$ with initial data: $p=1.01$, $a=0.01$.}
	\label{fig:figaa}
\end{figure}

Building on the previous discussion of the behavior of the field $ \phi $ around the critical value $ p_s $, we delve deeper into the nature of this transition, which reveals itself as a dynamical critical phenomenon — a phenomenon we refer to as a ``flip''. As the parameter $p$ approaches the critical point $ p_s $, the stable values of $ \phi $ exhibit a sign inversion on either side of the critical point. Specifically, this behavior can be captured by the relation:
\begin{equation}
	\lim_{p \to p_s^-} \phi(p) = -\lim_{p \to p_s^+} \phi(p).
	\label{eq:sign_inversion}
\end{equation}
A key aspect of this transition is that it is not exclusive to the variation along parameter $ p $; similar critical behavior is observed in other cases such as varying the charge $ Q $ and coupling parameter $ \beta $. As demonstrated in Fig. \ref{fig:fi} through Fig. \ref{fig:figaa}, this flip phenomenon is robust across different parameter regimes. Particularly in the phase diagram involving $ Q $, we observe an interesting feature: the occurrence of a ``double flip,'' where the system undergoes two successive sign changes as varying $ Q $. This behavior suggests a more intricate structure in the system's parameter space, hinting at the presence of multiple critical points and more complex forms of dynamical criticality. These results suggest that the gravitational backreaction system exhibits rich and nonlinear features, with the field $ \phi $ acting as a sensitive probe of the underlying critical phenomena.

The continuity of $|\phi_h|$ indicates that this flipping phenomenon results in an abrupt change in the sign of the scalar field without altering its magnitude. Given that our system possesses $Z_2$ symmetry, the spacetime geometry does not undergo a significant change. Consequently, this critical phenomenon cannot be discerned through the static characteristics of the spacetime geometry or the magnitude of scalar field $\phi$. Nevertheless, the dynamical behavior of the system allows us to capture this critical phenomenon effectively. 

\begin{figure}[H]
	\centering
	\includegraphics[width=0.48\linewidth]{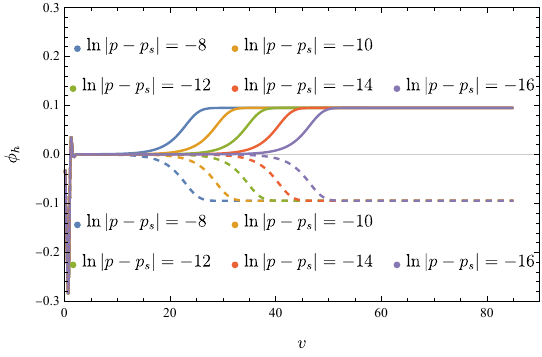}
	\includegraphics[width=0.48\textwidth]{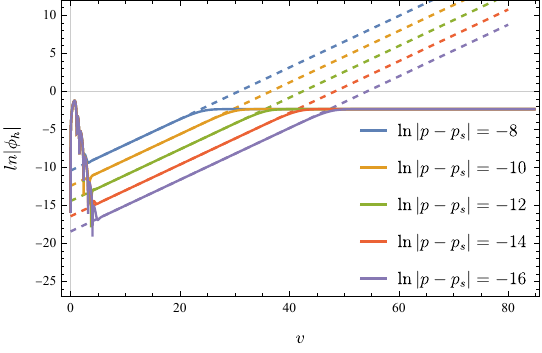}
	\caption{Evolution of $\phi_{h}$ and $\ln |\phi_{h}|$ over time, with initial data: $\beta=40$, $ Q = 0.6 $, $ a = 0.01 $, $p_{s}$ represents the flip point while $p_{s}=1.0158$. In the left panel, solid lines represent $p < p_s$ and dashed lines represent $p > p_s$. }
	\label{evolution_of_different_fi_flip}
\end{figure}

Near the flip points $p_{s}$, the relaxation dynamics are presented in Fig. \ref{evolution_of_different_fi_flip}, which illustrates the time evolution of $ \phi $ in the vicinity of the flip point. Specifically, numerical simulations demonstrate that the behavior of $ \phi $ is well-approximated by an exponential form:
\begin{equation}
	\phi(v) = e^{k_{\phi} v - b_{\phi}},
	\label{phi_rex}
\end{equation}
where $ k_{\phi} $ and $ b_{\phi} $ are fitting parameters. From this expression, we define the relaxation time as $ \tau_\phi = \frac{b_\phi}{k_\phi} $, which serves as a characterization of how long the system takes to relax the initial fluctuations. Importantly, the relaxation time depends sensitively on how close the system parameters are to the flip point. As the parameters approach their critical values, $\tau_\phi$ increases exponentially, reflecting the critical slowing down of the system's dynamics.

While the scalar field undergoes a sign inversion during the phase transition, it is crucial to note that, near the critical points, the metric and Maxwell fields exhibit a different but related behavior. Specifically, during the evolution near the flip point, as $ \phi $ approaches zero, the evolution of metric and Maxwell fields gradually change towards configurations that are close to the AdS-BI solution. This convergence suggests that the AdS-BI solution plays the role of the metastable state at the dynamical critical point.

In our prior investigations, the flip phenomenon observed in bald AdS-BI black holes has been fundamentally tied to the system's dynamics near an unstable critical point \cite{Wu:2024}. By introducing a small perturbation to this critical configuration, the scalar field evolves toward one of two distinct stable states, determined by the sign of the perturbation. This bifurcation arises from an inherent symmetry in the governing equations, leading to either positive or negative scalar hair configurations. The phenomenon underscores the system's sensitive dependence on initial conditions as it approaches the point of instability.

The relaxation dynamics associated with the flip are governed by a logarithmic scaling between the relaxation time and the initial perturbation amplitude. This behavior is analytically captured through quasi-normal mode (QNM) analysis, which identifies the dominant mode as the primary driver of the system's evolution. The perturbation $\delta\phi(v)$ evolves as  
\begin{equation}
	\delta\phi(v) \sim p e^{\omega_I v} = e^{\ln |p| + \omega_I v}
	\label{qnm}
\end{equation}
Consequently, the time required for the perturbation to amplify exhibits a logarithmic dependence on the perturbation magnitude. This highlights the intricate interplay between critical solutions, the stability of scalar hair within the black hole framework, and the universal dynamical features that emerge from the interaction of critical stability and perturbative evolution in black hole physics.

\begin{figure}[H]
	\centering
	\includegraphics[height=0.3\textwidth]{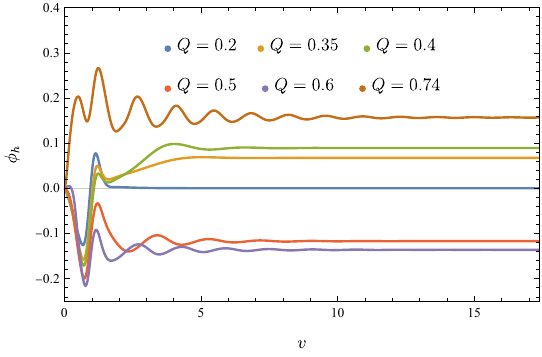}
	\includegraphics[height=0.3\textwidth]{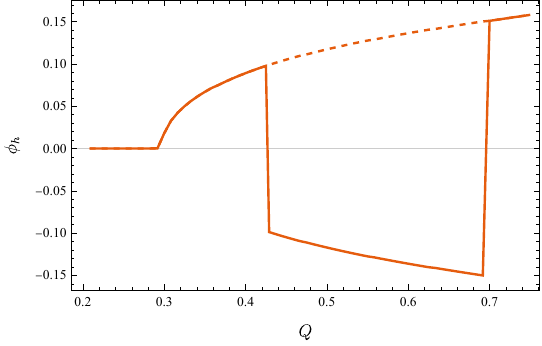}
	\caption{
		The left figure shows the evolution of the $\phi_{h}$ in systems with different $Q$ over time, while the right figure depicts the relationship between $Q$ and the final stable $\phi_{h}$ , with initial data: $p=0.64$, $\beta=150,\,a=0.01$.}
	\label{different_sclar_field}
\end{figure}

Additionally, Fig. \ref{different_sclar_field} provides further insights into the time evolution of the scalar field across systems with varying charge $Q$. The left panel shows the time evolution of the scalar field for different charge, while the right panel displays their corresponding phase mappings for the final value of scalar field. It is evident that as the charge increases, the scalar field undergoes more pronounced oscillations. These oscillations reflect the activation of distinct modes of the scalar field, illustrating the complex dynamic of the system. 

To further explore the physics in the vicinity of the dynamical critical point, it is crucial to explore the broader implications of this critical phenomenon within the context of AdS/CFT correspondence. In the next section, we shift our focus to HEE, which serves as a natural extension to probe the critical phenomena we have discussed.

\section{Introduction to Covariant Holographic Entanglement Entropy}
\label{sec:hee_intro}

A fundamental characteristic that makes quantum systems distinct from classical systems is entanglement, a phenomenon that can be assessed using various information-related metrics. HEE stands out as a well-established measure for quantifying entanglement, particularly in the context of pure states. In a quantum system described by a density matrix $\rho$, the entanglement of a subregion $A$ is defined as the von Neumann entropy of the reduced density matrix $\rho_A$:
\begin{equation}
	S_A(|\psi\rangle)=-\text{Tr}[\rho_A\text{log}(\rho_A)],\qquad \rho_A=\text{Tr}_B(|\psi\rangle\langle\psi|).
\end{equation}

In holography, the Ryu-Takayanagi (RT) prescription relates EE in a static, time-independent quantum field theory to the geometry of its dual AdS spacetime. Specifically, the holographic entanglement entropy (HEE) for a boundary subregion $A$ is proportional to the area of a minimal surface $\gamma_A$ in the bulk, whose boundary coincides with $\partial A$:
\begin{equation}
	S_A = \frac{\text{Area}(\gamma_A)}{4 G_N},
\end{equation}
where $ \gamma_A $ is the minimal surface whose boundary coincides with the boundary of $ A $ and $ G_N $ is Newton's gravitational constant. This remarkable formula not only provides insights into quantum gravity but also offers a powerful tool for studying quantum entanglement in strongly coupled systems.

To extend the concept of HEE to dynamical settings where the spacetime geometry and the entanglement structure evolve over time, the Hubeny-Rangamani-Takayanagi (HRT) prescription provides a covariant generalization of the RT formula \cite{Hubeny:2007xt}. This framework is particularly well-suited to handle time-dependent spacetimes, which naturally arise in contexts such as black hole formation, quantum quenches, or cosmological scenarios, where the entanglement entropy changes dynamically.

According to the HRT prescription, the entanglement entropy associated with a boundary subregion $\mathcal{A}$ is given by the following expression:
\begin{equation}
	S_\mathcal{A} = \frac{\text{Area}(\Gamma_\mathcal{A})}{4G_N^{(d+1)}},
\end{equation}
where $\Gamma_\mathcal{A}$ is an extremal surface in the bulk spacetime. This surface is defined such that its boundary coincides with the boundary of the subregion $\mathcal{A}$, i.e., $\partial \Gamma_\mathcal{A} = \partial \mathcal{A}$. The extremal condition requires that the area of $\Gamma_\mathcal{A}$ is stationary with respect to small perturbations, making it a saddle point of the area functional. In contrast to the RT formula, where the surface is minimal and the spacetime is static, the HRT prescription allows for extremal surfaces in fully dynamical, time-dependent bulk geometries.

The extremality condition of the surface $\Gamma_\mathcal{A}$ is central to the HRT proposal. It ensures that the surface has vanishing null expansions in both directions orthogonal to the surface, meaning that $\Gamma_\mathcal{A}$ is neither expanding nor contracting in these directions. This requirement allows the HRT prescription to capture the full causal and geometric structure of the bulk spacetime, making it valid in non-static settings where horizons, shocks, or other dynamical features are present.

In practical terms, when the bulk spacetime is described using advanced or retarded time coordinate systems, such as Eddington-Finkelstein coordinates, the extremal surface $\Gamma_\mathcal{A}$ is determined by solving the equations of extremality while ensuring that the surface remains anchored to the boundary of the subregion $\mathcal{A}$ on the specified time slice corresponding to $v$. This approach allows the HRT prescription to account for the intrinsic time dependence of the bulk and, by extension, the time evolution of entanglement entropy in the boundary quantum field theory.

For our numerical solutions, we use the following metric in ingoing Eddington-Finkelstein coordinates. As shown in \eqref{eq:ansatz}. To address computational challenges associated with the infinite nature of the space, we redefine the radial coordinate $r$ as $z = r_h/r$, leading to the reformulated metric:
\begin{equation}
	ds^2 = -\alpha(v,z) dv^2 - \frac{2}{z^2} dvdz + \zeta(v,z)^2(d\theta^2 + \sin^2 \theta d\varphi^2)
\end{equation}
Furthermore, we parameterize the minimal surface using $\vartheta$, expressing $v$, $z$, and $\theta$ as functions of $\vartheta$, where $\vartheta$ is chosen to be constrainted by
\begin{equation}
	z(\vartheta)=r_{h} \theta(\vartheta)  \tan(\vartheta)=\theta(\vartheta) \tan(\vartheta).
	\label{eq:difine}
\end{equation}
Substituting these expressions into the metric, we obtain:
\begin{equation}
	ds^2 = 
	\left[-\alpha(v(\vartheta), z(\vartheta)) v'(\vartheta)^2
		- \frac{2}{z(\vartheta)^2} z'(\vartheta) v'(\vartheta) 
		+ \zeta(v(\vartheta), z(\vartheta))^2 \theta'(\vartheta)^2 
		\right] d\vartheta^2 
	+ \sin^2 \theta(\vartheta) \, d\varphi^2.
\end{equation}
The area of the minimal surface can be expressed as a Lagrangian system:
\begin{equation}
	\label{integrateneed}
	S = 2\pi \int \mathcal{L} \, d\vartheta
\end{equation}
where the Lagrangian $ \mathcal{L} $ is given by:
\begin{equation}
	\mathcal{L} = \sqrt{-\alpha(v,z) v'^{2} - \frac{2}{z^2} z' v' + \zeta(v,z)^{2} \theta'^{2}} \zeta(v,z) \sin{\theta}
\end{equation}
To find the minimal surface, we need to solve the Euler-Lagrange equations:
\begin{equation}
	\frac{d}{d\vartheta}\left(\frac{\partial L}{\partial q'(\vartheta)}\right) - \frac{\partial L}{\partial q(\vartheta)} = 0, \quad q = \theta, z, v
	\label{eq:Lag}
\end{equation}
In the search for minimal surfaces in $4$-dimensional spherically symmetric spacetime, there exist two types of configurations: disk and spherical minimal surfaces, each subject to different boundary conditions. For computational convenience, we have selected the disk minimal surface configuration for our analysis. Accordingly, we can define the boundary conditions using the radius  $\mathfrak{R} = \theta(0)$, as specified in \eqref{eq:bou} \cite{{Ling:2019tbi}}.

\begin{equation}\label{eq:bou}
	z(0) = z'\left(\frac{\pi}{2}\right) = 0, \quad v(0) = v_b, \quad v'\left(\frac{\pi}{2}\right) = 0, \quad \theta(0) = \mathfrak{R}, \quad \theta'\left(\frac{\pi}{2}\right) = 0
\end{equation}
Here, $ v_b $ represents the initial time at which the evolution begins.

Having obtained the time evolution results for the background metric solutions, we utilize the numerical functions $\alpha$ and $\zeta$ to compute the minimal surface. The minimal surface in spacetime is determined by solving a system of equations composed of three Lagrange equations \eqref{eq:Lag}, of which two are independent(the rank of equations \eqref{eq:Lag} is 2), in addition to the line parameterization equation \eqref{eq:difine} and the boundary conditions \eqref{eq:bou}. However, in practical numerical computations, resolving only two of the equations from \eqref{eq:Lag} along with \eqref{eq:difine} often leads to various numerical difficulties. To address this issue, we include all three equations from \eqref{eq:Lag} as part of the system to be solved, along with \eqref{eq:difine}. 

After discretizing the minimum surface using Chebyshev-Lobatto collocation, we employ the least squares iterative method to minimize the combined error across all equations, ensuring that the error remains below the threshold of $10^{-7}$. This approach effectively mitigates the apparent overdetermination of the system and enhances the optimization of the numerical solution process, ultimately enabling the successful determination of the minimal surface in spacetime.

\begin{figure}[H]
	\centering
	\includegraphics[width=0.6\textwidth]{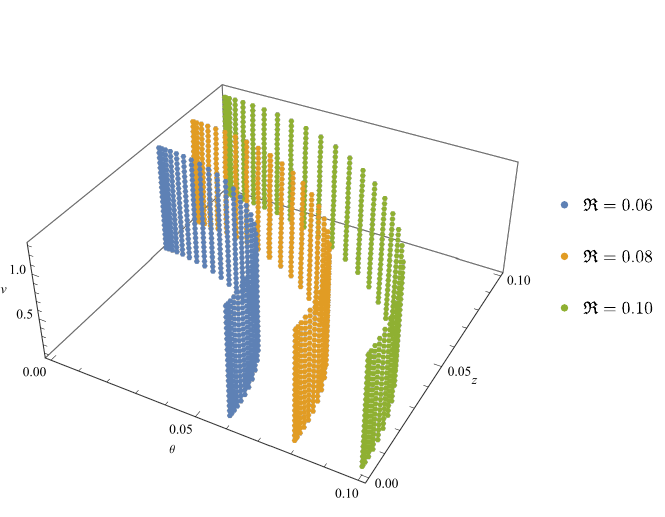}
	\caption{
		Diagram of four-dimensional minimal surfaces at different times $v_{b}$, with initial data: $p=0.89673$, $\beta=60$, $ Q = 0.6 $, $ a = 0.01 $.}
	\label{fig:4_demintion}
\end{figure}

Building upon the above approach, Fig. \ref{fig:4_demintion} presents a diagram of the minimal surfaces at various times slice. Given the relatively modest changes in our metric, the time variations of the minimal surfaces are negligible in visual inspection. However, rigorous numerical comparisons reveal differences on the order of $10^{-2}$ between these surfaces. Despite these seemingly small alterations, our numerics proves robust in capturing effective time-dependent changes in the covariant HEE. This sensitivity underscores the efficacy of our technique in extracting meaningful results even from apparently minor geometric variations, a point we will elaborate on in subsequent sections.

After obtaining the minimal surface, the minimal surface area can be calculated by integrating along the minimal surface. However, it is important to emphasize that this integral will diverge when $\vartheta=0$. Therefore, we will follow these steps for calculation:

\begin{enumerate}
	\item Obtain the numerical data of minimal surface at a certain time and interpolate it.
	\item Substitute the interpolation function into the integral \eqref{integrateneed} and expand the integrand at $\vartheta = 0$ to extract the divergent terms.
	\item Subtract the divergent terms from the integrand for numerical integration.
	\item Perform an analytical integral on the divergent terms, substitute $\vartheta = \frac{\pi}{2}$ into the integral result, and then incorporate this contribution into the result from step 3.
\end{enumerate}

Before exploring dynamical evolutions, we first focus on the case of stabilized solutions. This allows us to clearly identify how the BI parameter $ a $ affects the behavior of HEE compared to standard electromagnetic scenarios. By examining these static solutions, we can establish a baseline understanding of the system's behavior, setting the stage for a deeper analysis of the dynamics that follow. Let us now turn to the static holographic entanglement entropy in the EBIS framework.

\section{Probing Phase Transitions with Static HEE}\label{sec:hee_static}

The HEE has emerged as a key concept in the study of quantum many-body systems, offering deep insights into their complex behavior. As a fundamental measure of quantum correlations, the entanglement entropy not only quantifies the degree of entanglement between a given subregion and its complement but also serves as a probe of the system's thermodynamic properties. In this section, we explore the final-state HEE of the EBIS system, examining its dependence on the initial wave packet amplitude $ p $, the system's charge $ Q $, and the BI factor $ a $.

\begin{figure}
	\subfloat[]{\includegraphics[width=0.48\textwidth]{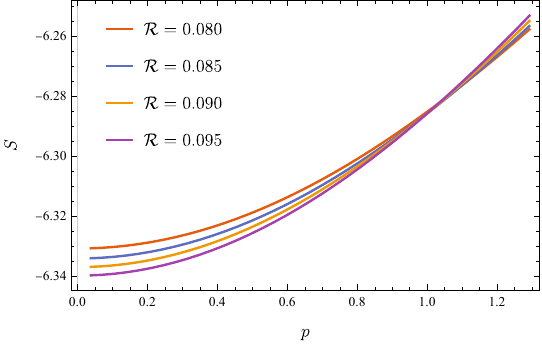}\label{fig:pandS}}
	\subfloat[]{\includegraphics[width=0.48\textwidth]{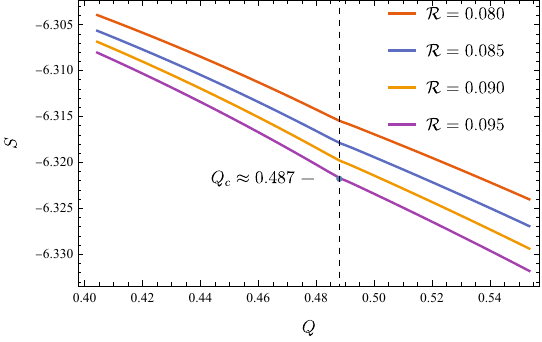}\label{fig:QandS}}\\
	\subfloat[]{\includegraphics[width=0.48\textwidth]{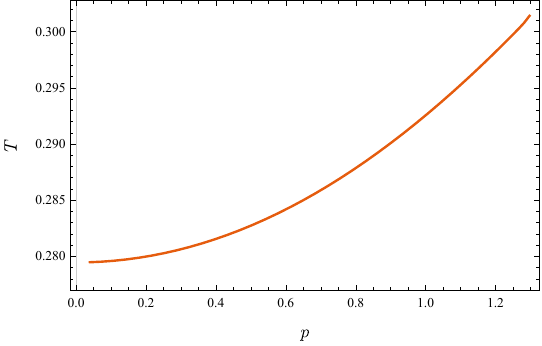}\label{fig:pandT}}
	\subfloat[]{\includegraphics[width=0.48\textwidth]{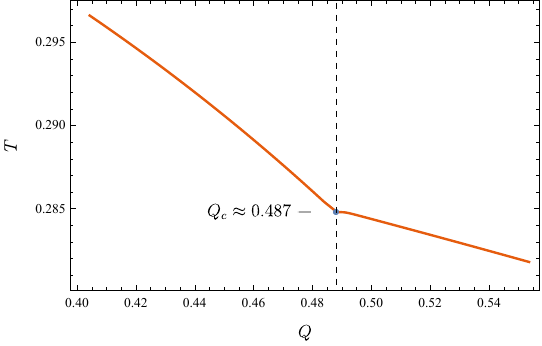}\label{fig:QandT}}\\
	\subfloat[]{\includegraphics[width=0.48\textwidth]{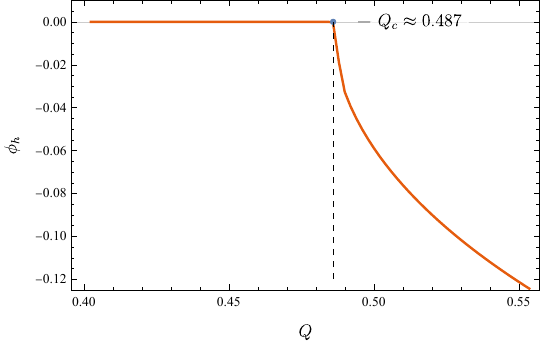}\label{fig:Qandphi}}
	\captionsetup{justification=raggedright,singlelinecheck=false}
	\caption{
		Panel (a) and (c) show the HEE and temperature $T$ versus $p$, with initial data: $\beta=40$, $Q=0.6$, $a=0.01$. Panel (b), (d), and (e) show the HEE, temperature $T$, and final value of $\phi_h$ versus $Q$, with initial data: $p=0.005$, $\beta=40$, $a=0.01$.
	}
	\label{fig:static_rep}
\end{figure}

The numerical results indicate that the final-state HEE monotonically increases with the wave packet amplitude $ p $, as shown in the Fig. \ref{fig:pandS}. Similarly, the temperature of the final-state black hole in the  also increases with $ p $, as depicted in Fig. \ref{fig:pandT}. This monotonic relationship between the HEE and black hole temperature has been verified and explicitly revealed in \cite{Liu:2023rhd}, establishing a clear connection between the entanglement entropy and the thermal entropy of the system.

On the other hand, as shown in Fig. \ref{fig:QandS}, the final-state HEE decreases monotonically as the black hole charge increases. This behavior is closely associated with the black hole temperature, which also decreases with increasing charge, as depicted in Fig. \ref{fig:QandT}. These results once again manifest the established relationship between the HEE and the system's temperature, as previously discussed. 
Additionally, the rate at which the HEE decreases is significantly slower than the rate of temperature decrease, suggesting that a larger black hole charge introduces additional entangled degrees of freedom to counterbalance the effects of the cooling system. Notably, as $Q$ approaches a critical value $Q_c$ (indicated by the dashed line in  Fig. \ref{fig:QandS} and \ref{fig:QandT}), a discontinuity emerges in the first derivative of both the HEE and the black hole temperature with respect to $Q$. This discontinuity signals the occurrence of a phase transition between the scalarized phase and the normal phase. This critical point corresponds to the phase transition identified in the broader phase diagram of the system, as shown in Fig. \ref{fig:Qandphi}.

\begin{figure}[H]
	\centering
	\includegraphics[width=0.6\textwidth]{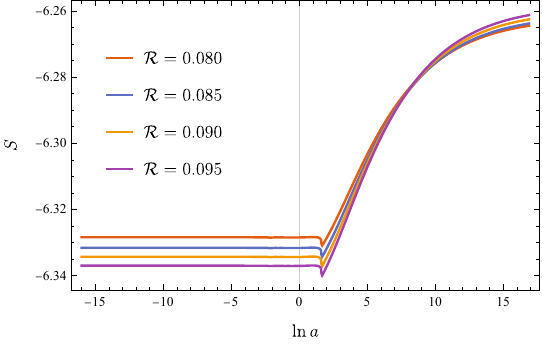}
	\caption{The relationship between $ \ln(a) $ and the HEE, with initial date: $p=0.64$, $\beta=40$, $Q=0.7$}
	\label{fig:entanglement_entropy}
\end{figure}

To examine the influence of the nonlinear corrections to the electromagnetic field on the HEE, we analyze the relationship between the BI factor, denoted as $ a $, and the HEE, $ S $, as depicted in Fig. \ref{fig:entanglement_entropy}. In the left portion of the plot, the HEE remains constant as $ a $ increases, suggesting that the system's properties in this region resemble those of the EMS limit and are relatively insensitive to the BI corrections. However, as $ a $ continues to grow, the HEE begins to increase significantly, signaling a transition to a regime where the system exhibits characteristics more akin to the ES limit, with the nonlinear corrections playing a more substantial role. Notably, near $\ln(a) \approx 1.6$, the HEE shows a striking non-monotonic behavior within a narrow range of $ a $, at this critical point the system undergoes a transition from the EBIS black hole to the AdS-BH black hole. As the nonlinear corrections become more prominent, the enhanced self-interactions of the BI field result in a more entangled system, leading to a higher HEE.

The results presented in this section provide a comprehensive overview of the final-state HEE in the EBIS system, highlighting the intricate and significant role of the nonlinear electromagnetic corrections in shaping the system's entanglement structure. Next, we will delve into the dynamic evolution of the HEE, exploring how the HEE responds to the dynamical processes in the EBIS system.

\section{HEE as a Diagnostic of Scalar Field Dynamics and Critical Phenomena}\label{sec:hee_dynamic}

HEE has emerged as a powerful tool for probing various dynamical processes in condensed matter systems, prompting considerable interest in its potential to elucidate dynamical evolution and critical phenomena \cite{Liu:2023rhd}. In this section, we examine the time evolution of HEE in spontaneous scalarization, focusing on the specific characteristics induced by such evolution. We first explore how the dynamical critical phenomenon can be captured by HEE. Further, we investigate the time evolution of HEE in response to scalar field dynamics, which exert distinct influences on the system. This analysis highlights the capacity of HEE to reflect both dynamical critical phenomena and spontaneous scalarization process. Lastly, we analyze the role of the BI factor, $a$, in shaping the dynamic properties of HEE.

\begin{figure}
	\centering
	\includegraphics[width=0.6\linewidth]{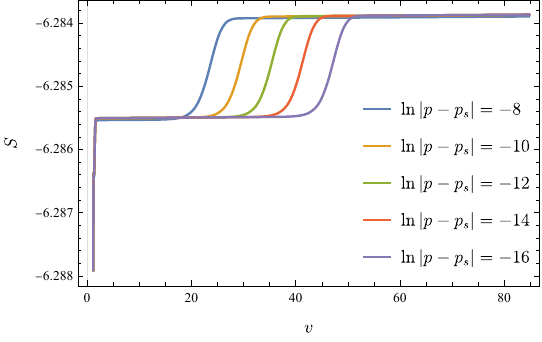}
	\captionsetup{justification=raggedright,singlelinecheck=false}
	\caption{The time evolution of HEE, with $ Q = 0.6 $, $ a = 0.01 $, $\beta=40$, $\mathfrak{R}=0.08$. The selection of different widths $\mathfrak{R}$ will lead to changes in the absolute value of the HEE, but it does not affect the HEE's ability to capture dynamic phenomena. Therefore, we will illustrate this with $\mathfrak{R} = 0.08$ as an example.}
	\label{HEE_filp_2}
\end{figure}

To explore the manifestation of dynamical critical phenomena in HEE, we analyzed its time evolution near the flip point. Our findings demonstrate a relaxation behavior reminiscent of that observed for scalar fields near the flip point, as illustrated in the right penal of Fig. \ref{HEE_filp_2}. Here, $p_s$ denotes the flip-point value of the initial amplitude of the wave packet. Remarkably, HEE exhibits a relaxation pattern strikingly similar to that of the scalar field, whose behavior is depicted in Fig. \ref{evolution_of_different_fi_flip}. This similarity suggests that the relaxation dynamics of HEE can be characterized analogously to the framework established for the relaxation of scalar fields,
\begin{equation}
	\label{discrabe}
	\begin{aligned}
		S = S_0 + \delta S \approx S_0 + e^{k_{EE}v - b_{EE}},
	\end{aligned}
\end{equation}
where $k_{EE}$ and $e^{-b_{EE}}$ represent the growth rate and the initial amplitude of $\delta S$, respectively. To further quantify the dynamics, we define the relaxation time of the dynamic HEE as $\tau_{EE} \equiv \frac{b_{EE}}{k_{EE}}$, which describes the relaxation process. Notably, the relaxation behavior emerges as a distinguishing characteristic of the HEE dynamics near the flip point. Moreover, while the scalar field $\phi$ may undergo sign flips, the HEE, which depends on the metric, does not directly reveal these sign changes since the equations of motion show that the metric interacts with $\phi^2$, meaning only the magnitude affects the entanglement structure.

\begin{figure}[H]
	\centering
	\includegraphics[width=0.48\linewidth]{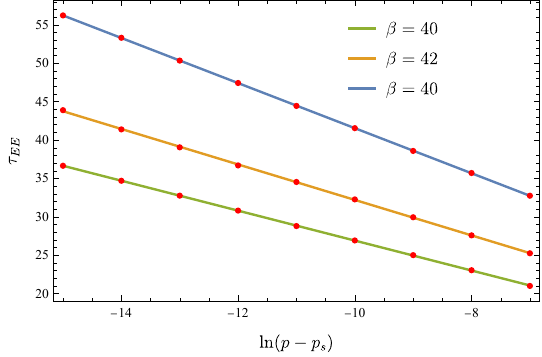}
	\includegraphics[width=0.48\textwidth]{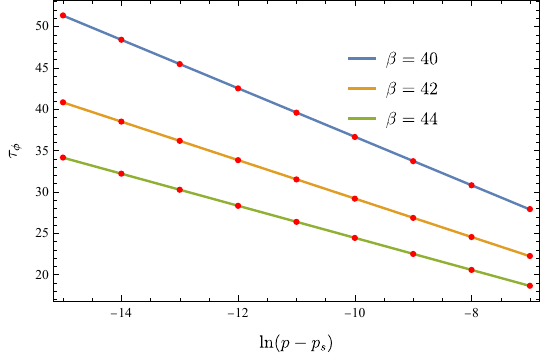}
	\caption{The relationship between the perturbation strength $\ln |p - p_s|$ and the relaxation time of the HEE and $\phi_{h}$, for parameters $ Q = 0.6 $, $ a = 0.01 $, $\mathfrak{R}=0.08$. }
	\label{linear_HEE}
\end{figure}

In Fig. \ref{linear_HEE}, we present the relationship between the logarithmic parameter distance, $\ln \Delta p = \ln |p - p_s|$, and the relaxation time of HEE, $\tau_{EE}$, and the scalar field, $\tau_{\phi}$. The results demonstrate a clear proportional relationship between the relaxation time $\tau$ and $\ln |p - p_s|$. Specifically, the relaxation time of the scalar field shows a linear dependence on $\ln \Delta p$, which can be comprehensively understood within the framework of QNM, as described in Eq. \eqref{qnm}. Furthermore, the relaxation behavior of HEE suggests that the system's metric undergoes a relaxation process that mirrors the dynamics of the scalar field, reinforcing a connection between scalar field dynamics and the response of the spacetime geometry. 

It is important to emphasize that the parameters $ k_\phi $ and $ b_\phi $ (and consequently $ \tau_\phi $) governing the dynamics of the scalar field, as well as $ k_{EE} $ and $ b_{EE} $ (and thereby $ \tau_{EE} $) describing the behavior of HEE, are intrinsically dependent on system parameters such as coupling constant $ \beta $. As $ \beta $ varies, these relaxation parameters shift independently, leading to corresponding changes in $ k_\phi $ and $ k_{EE} $. While this dependence is anticipated for the scalar field based on the well-known perturbation analysis, the relaxation dynamics of HEE are less straightforward. Since HEE is fundamentally linked to the geometry of the spacetime, its evolution effectively encodes the relaxation of the underlying spacetime metric. This suggests that $\delta S \sim \delta g \sim \phi^\alpha$ in the perturbation regime, where $\alpha$ denotes an undetermined power. Consequently, the relationship between the timescales $ \tau_\phi $ and $ \tau_{EE} $ may exhibit non-trivial characteristics.

To further characterize the relationship between $\tau_{EE}$ and $\tau_{\phi}$, we define a constant $K_{\#}$ as follows:
\begin{equation}
	\label{ks_definition}
	K_{\#} \equiv \frac{\partial \tau_{\#}}{\partial \ln|p-p_s|},
\end{equation}
where $\#$ represents either the subscript of scalar field or the HEE. Interestingly, our numerical results reveal that $K_{EE}$ is numerically equivalent to $K_{\phi}$. Although $K_{EE}$ and $K_{\phi}$ are derived from the relaxation parameters of distinct fields—the spacetime metric (associated with HEE) and the scalar field, respectively—they encapsulate the same underlying physical mechanism. This equivalence suggests a deep connection between the relaxation dynamics of HEE and scalar field, both governed by the same fundamental processes.

To further elucidate this observation, we rewrite the relaxation dynamics of HEE and the scalar field as follows:
\begin{equation}\label{discrabe2}
	\begin{aligned}
		S(v) = S_{0} + e^{k_{EE}v - k_{EE}\tau_{EE}}, \quad \phi_h(v) = e^{k_\phi v - k_\phi \tau_\phi}.
	\end{aligned}
\end{equation}
Here, $ S_0 $ represents the initial value of the HEE, which is non-zero. The time required for the system to reach equilibrium can then be expressed as:
\begin{equation}\label{discrabe3}
	\begin{aligned}
		v'_{EE} = \frac{\ln(S' - S_{0})}{k_{EE}} + \tau_{EE}, \quad v'_\phi = \frac{\ln \phi'_h}{k_\phi} + \tau_\phi.
	\end{aligned}
\end{equation}
In these expressions, $ v'_\phi $ and $ v'_{EE} $ denote the time scales required for the scalar field and the HEE to reach their final-state value $ \phi'_h $ and $ S' $. Notably, the scalar field and the HEE evolve synchronously, reaching their final states simultaneously during the relaxation process. This observation implies that $ v'_{EE} = v'_\phi $. From this equality, a significant conclusion follows: near the dynamical critical point of the system (the flip point), the variation in relaxation times satisfies $ \delta \tau_{EE} = \delta \tau_{\phi} $. Here, $\delta \tau_{EE}$ and $\delta \tau_{\phi}$ are defined as the changes in relaxation times corresponding to the same infinitesimal variation in the logarithmic parameter distance, $\delta \ln |p - p_s|$, where $p_s$ is the critical parameter value at the flip point. This relationship holds because near the flip point, variations in $p$ lead to negligible changes in the final values $S'$ and $\phi'_h$, effectively making $S'-S_0$ approximately constant in this regime.  Consequently, the equivalence $ K_{EE} = K_{\phi} $ holds, as $ K_{EE} $ and $ K_{\phi} $ are both defined as the rate of change of the relaxation time $ \tau $ with respect to the logarithm of the parameter distance, $ \ln \Delta p $.

This result demonstrates that, while the intermediate stages of relaxation may differ in their specific details, the overall characteristic timescale of relaxation is the same. Notably, the equality $K_{EE} = K_{\phi}$ highlights the profound connection between the dynamics of the scalar field and the behavior of the HEE in the vicinity of the critical point. This analysis suggests that the relaxation dynamics of the HEE closely mirror those of the scalar field. Moreover, the characteristic relaxation behavior of the HEE aligns closely with that of the scalar field, implying that the dynamical properties of the HEE can serve as a diagnostic tool to identify the critical point in the system's evolution.

\begin{figure}[H]
	\centering
	\includegraphics[height=0.3\textwidth]{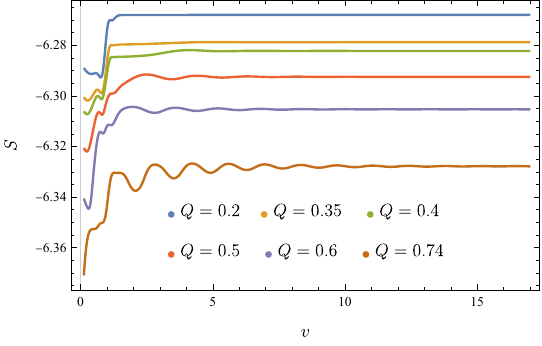}
	\includegraphics[height=0.3\textwidth]{figures/figqep.pdf}
	\caption{
		The left figure shows the evolution of HEE  with different $Q$ over time, while the right figure depicts the evolution of $\phi_{h}$ , with initial data: $p=0.64$, $\beta=150$, $\mathfrak{R}=0.08$, $a=0.01$. }
	\label{heeflipo}
\end{figure}

In the previous discussion, we mentioned that altering the parameters of the system can change the oscillatory behavior of the scalar field. To further investigate whether HEE can characterize the dynamic of the system, we calculated the time evolution of the HEE for different charges. The left panel of Fig. \ref{heeflipo} shows the time evolution of the HEE, while the right panel displays the time evolution of the scalar field under the same parameters setting.

The numerical results indicate that as the charge increases, the oscillation of the scalar field exhibits higher frequency, suggesting that the system undergoes more significant changes before reaching its final state. Similarly, with the increase in charge, the time evolution of HEE reveals more rapid oscillations, demonstrating behavior akin to that of the scalar field. This indicates that the dynamical behavior of the scalar field can be effectively captured by HEE. Thus, HEE serves as a valuable indicator of the dynamical evolution process, reflecting a substantial amount of information regarding the dynamics of the system's evolution.

\begin{figure}[H]
	\centering
	\includegraphics[width=0.6\textwidth]{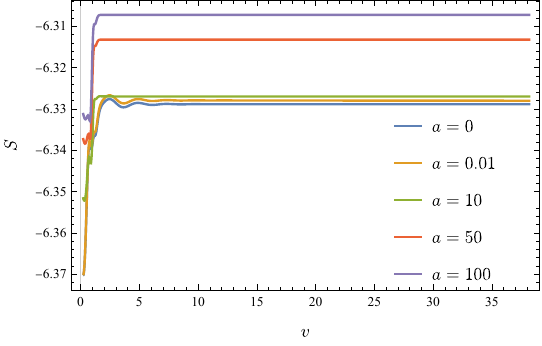}
	\caption{Time evolution of HEE, with parameters set to $Q=0.7$, $p=0.64$, $\beta=40$, $\mathfrak{R}=0.08$.}
	\label{fig:bi_dynamic}
\end{figure}

In Fig. \ref{fig:bi_dynamic}, we present the time evolution of HEE as the BI factor $ a $ varies. It can be observed that with the increasing BI factor $ a $, the overall entanglement of the system exhibits a rising trend. Furthermore, when the system approaches the EMS phase (with small values of $ a $), it shows higher frequency and amplitude of oscillation. 
This behavior is further corroborated in Fig.~\ref{heeflipo}, where we demonstrate that increasing the system's charge enhances the oscillatory dynamics. On the other hand, the BI factor $ a $, being a nonlinear correction term to the electromagnetic field, limits the maximum strength of the electromagnetic field in this system. When $ a $ is large, the system tends to resemble the ES system, where the electromagnetic field ceases to play a significant role, leading to a natural reduction in the system's oscillatory behavior. This behavior is consistent with the trends observed in the HEE presented in Fig. \ref{fig:bi_dynamic}.

Up to this point, we have demonstrated that the dynamic behavior of HEE can serve as a valuable tool to characterize the evolution of the scalar field. Our numerical computations support this assertion, showing that dynamical evolution processes such as slow and rapid dissipation of scalar hair, as well as dynamical critical phenomena like flip phenomena, are manifested in the time evolution of HEE. Thus, we can conclude that HEE serves as a sensitive probe to capture the dynamics of evolution, offering a novel perspective on the evolution of the system and the processes of dynamical critical phenomena.

\section{Discussion}
\label{sec:discussion}

In this work, we have demonstrated that holographic entanglement entropy (HEE) serves as a powerful diagnostic tool for both static and dynamical critical phenomena in the Einstein-Born-Infeld-Scalar (EBIS) model. Our findings bridge the gap between equilibrium and nonequilibrium physics in holographic systems, offering new insights into the interplay between quantum information, holography, and gravitational dynamics.

HEE has been widely used to identify static critical phenomena in holographic systems, such as thermal and quantum phase transitions \cite{Liu:2020blk,ling2016characterization,ling2016holographic,Liu:2023rhd,Chowdhury:2024lqw,Barbosa:2024pyn,Yang:2023wuw,Liu:2021rks}. In the EBIS model, we found that HEE exhibits a discontinuity in its first derivative at the critical point separating scalarized black holes and AdS-Born-Infeld (AdS-BI) black holes. This discontinuity serves as a clear signature of the phase transition, reinforcing the utility of HEE as a diagnostic tool for static critical phenomena.

While entanglement entropy is well-known for capturing static criticality, its ability to probe dynamical critical phenomena has remained less explored. Our study addresses this gap by showing that HEE can capture the ``flip'' phenomenon, a dynamical criticality characterized by a sign inversion in the scalar field at a critical point. Near the flip point, HEE exhibits relaxation dynamics described by $\delta S \sim e^{k_{EE}v - b_{EE}}$, mirroring the exponential relaxation of the scalar field, $\phi_h \sim e^{k_\phi v - b_\phi}$. Both relaxation timescales follow the same logarithmic scaling, $\tau \propto \ln |p - p_s|$, where $p_s$ is the critical value. This scaling behavior is consistent with quasi-normal mode (QNM) analysis, suggesting a deep connection between the relaxation dynamics of HEE and the underlying gravitational dynamics.

Notably, the proportionality constants $K_{EE}$ and $K_\phi$, which describe the rate of change of relaxation times with respect to the logarithmic parameter distance, are numerically equivalent ($K_{EE} = K_\phi$). This equivalence underscores that the relaxation dynamics of HEE are governed by the same fundamental processes as those of the scalar field, highlighting HEE as a sensitive probe of dynamical criticality.

Our results provide a new perspective on the relationship between bulk dynamics and boundary physics in the AdS/CFT correspondence. The relaxation dynamics of HEE reflect the underlying gravitational dynamics, suggesting that quantum informational observables on the boundary can encode detailed information about bulk processes. This connection may have implications for the study of quantum gravity, where entanglement entropy is believed to play a central role in the emergence of spacetime geometry.

Our findings open up several promising directions for future exploration. One natural extension involves the investigation of planar black hole configurations, which could exhibit additional dynamical features and subtle quantum information signatures, such as entanglement wedge cross-sections (EWCS) and mutual information (MI) \cite{Liu:2020blk,Liu:2023rhd,Yang:2023wuw,Liu:2021rks,Liu:2019npm,Huang:2019zph,Chen:2021bjt,Li:2023edb,Chen:2024wpt,Cai:2017ihd}. These finer measures may provide deeper insights into how diverse geometric structures influence the evolution of entanglement and correlations at the boundary. Moreover, extending our framework to other gravitational theories, such as those incorporating higher-curvature corrections or alternative couplings, could reveal the robustness and universality of the relationships we observed. Such studies may enhance our understanding of the interplay between holography, quantum information measures, and gravitational dynamics, while also uncovering potential new principles governing critical phenomena and dynamical processes in more complex settings.

\section*{Acknowledgements}

Peng Liu would like to thank Yun-Ha Zha and Yi-Er Liu for their kind encouragement during this work. This work is supported by the Natural Science Foundation of China under Grant No. 12375048 and 12475054.

\end{document}